\documentclass[twocolumn,fleqn,10pt]{wlscirep}
\usepackage[utf8]{inputenc}
\usepackage[T1]{fontenc}
\usepackage{amssymb,amsmath,gensymb}
\usepackage{lipsum}

\usepackage[normalem]{ulem}
\usepackage{changes}
\usepackage{chngcntr}
\usepackage{epsfig}
\usepackage{amssymb}
\usepackage{soul}
\usepackage{color}
\usepackage{float}
\usepackage{textcomp}
\usepackage{amstext}
\usepackage{graphicx}
\usepackage{gensymb}
\usepackage{graphics}\usepackage{subfigure}\usepackage{longtable}\usepackage{pstricks}\usepackage{dcolumn}\usepackage{bm}
\usepackage{siunitx}
\usepackage{wasysym}
\usepackage[font=small,labelfont=bf]{caption}

\makeatletter
\newcommand*{\balancecolsandclearpage}{%
  \close@column@grid
  \cleardoublepage
  \twocolumngrid
}
\makeatother

\title{Separating the Density Wave Phase from Zero-Resistance Superconductivity in La$_4$Ni$_3$O$_{10}$}
\author[1,*]{P.~Puphal}
\author[1]{A.~Suthar}
\author[2]{B.~Wehinger}
\author[1]{V.~Sundaramurthy}
\author[1]{T.~Priessnitz}
\author[1]{M.~Knauft}
\author[1]{P.~Sosa-Lizama}
\author[1]{Y.-M.~Wu}
\author[1]{M.~Hepting}
\author[1]{J.~Nuss}
\author[1]{M.~Isobe}
\author[1]{P.~A.~van~Aken}
\author[1]{Y.~E.~Suyolcu}
\author[1]{B.~Keimer}
\author[1]{H.~Takagi}
\author[1,**]{P.~Reiss}
\affil[1]{Max Planck Institute for Solid State Research, Heisenbergstra{\ss}e 1, 70569 Stuttgart, Germany}
\affil[2]{European Synchrotron Radiation Facility, 71 Avenue des Martyrs, F-38043 Grenoble, France}
\affil[*]{puphal@fkf.mpg.de}
\affil[**]{p.reiss@fkf.mpg.de}

\begin{abstract}
Identifying the key ingredients for superconductivity under pressure in the Ruddlesden-Popper nickelates remains a formidable challenge, due to a complex interplay between structural variations, oxygen content and location, electronically ordered phases, and the need for high pressures to reach the superconducting phase. To disentangle these complex inter-dependencies, here we study a series of clean single crystals of trilayer La$_4$Ni$_3$O$_{10}$. We discover that enhanced biaxial, compressive pressures induce a zero-resistance state at a 60\% lower pressure compared to previous reports. At the same time, we completely separate the low-pressure density wave phase from superconductivity, effectively ruling out that soft, quantum critical fluctuations of the density wave order parameter act as the pairing glue in this system. Our results establish biaxial, in-plane pressures as a powerful tool for promoting superconductivity in nickelates, unifying the required pressure scales for superconductivity in different Ruddlesden-Popper structures, and providing a connection to the recently discovered thin film samples.
\end{abstract}

\begin{document}

\flushbottom
\maketitle

\thispagestyle{empty}

The observation of high-temperature superconductivity under pressures in single crystals of Ruddlesden-Popper nickelates has spurred an enormous effort by the scientific community to identify the necessary conditions for the superconducting pairing \cite{Sun2023,Wang2024,Zhang2024,Zhang2025,Li2026,Sakurai2026}. However, progress towards this goal has been held back by a formidable challenge to achieve a state featuring a robust zero resistance. In fact, most reports to date only report signatures of superconductivity, e.g. a field- and current-dependent downturn in the electrical resistivity, or a weak signature in DC and AC susceptibilities \cite{Zhou2025}. A plethora of difficulties for achieving robust superconducting phases have been identified, covering the full spectrum from sample growth conditions to details of the experimental setups employed: examples include the precise stacking order of the NiO$_2$ planes \cite{Wang2023,Puphal2023,Chen2024a,Li2024b,Huang2025}, oxygen deficiency or excess as well as the location of oxygen ions within the perovskite and/or rock salt layers \cite{Dong2024,Chen2025,Jiang2025a,Hu2026}, the crystal structure at ambient and high pressure conditions \cite{Sun2023,Wang2023a,Puphal2023,Shi2025a,Misawa2026,Reiss2026}, doping on the lanthanum or nickel sites \cite{Wang2024,Li2026}, the degree of (non-)hydrostaticity of the pressure environment \cite{Wang2023b,Hou2023,Mandyam2026}, and the presence or absence of electronic order at ambient pressure at low and high temperatures \cite{Liu2023,Chen2024,Khasanov2024,Shi2025a,Reiss2026}.

In order to disentangle this seemingly impenetrable web of correlated and/or conflicting ingredients, we focus on the high-pressure superconducting trilayer compound La$_4$Ni$_3$O$_{10}$ \cite{Zhu2023,Li2024,Sakakibara2024,Zhang2025a}. In contrast to bilayer La$_3$Ni$_2$O$_7$, the growth conditions of the trilayer phase are much more stable, which avoids the occurrence of stacking variations or phase intergrowth, and therefore provides a more reliable platform for investigating the electronic properties \cite{Zhang2020a,Reiss2026}. Our central approach is to control the high-pressure environment and its non-hydrostatic components explicitly \cite{Dunstan1989a}. We find that enhancing the in-plane, biaxial pressure leads to a robust and reproducible zero-resistance superconducting phase in La$_4$Ni$_3$O$_{10}$, at 60\% lower pressures compared to previous studies. Moreover, we completely separate the low-pressure density wave phase from high-pressure superconductivity \cite{Zhang2020,Shi2025}. Our report unifies the pressure scales required for achieving superconductivity between the bilayer and trilayer phases, and provides a link to the epitaxially strained thin film samples \cite{Ko2025,Zhang2026}. Moreover, we find that previous transport studies on nickelates most likely experienced an unreported strong out-of-plane pressure component. Not only does this component appear to be detrimental to the formation of superconductivity, but it is also absent during typical structural and magnetisation studies, challenging the interpretation of previous comparative reports. Consequently, our report establishes that biaxial, compressive pressures are a powerful tool for tuning the electronic phase diagram and high-pressure superconductivity in Ruddlesden-Popper nickelates.

\begin{figure}[htb]
\includegraphics[width=75mm]{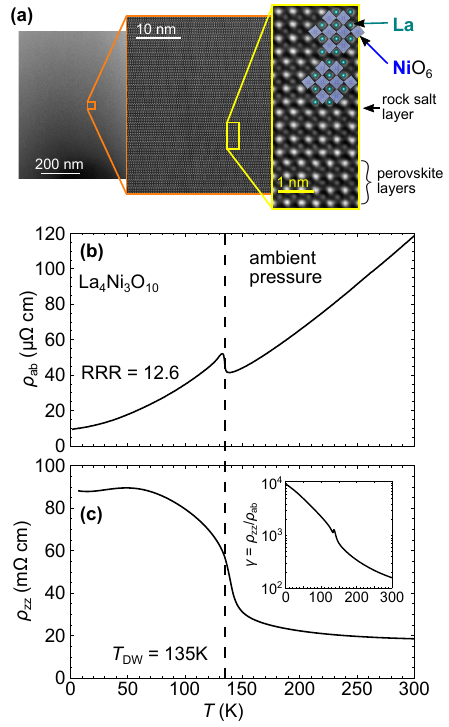}
\caption{Ambient pressure properties of La$_4$Ni$_3$O$_{10}$. 
(a) Wide view/close up high-angle annular dark-field scanning transmission electron microscopy of La$_4$Ni$_3$O$_{10-x}$ single crystal. The overlay show the monoclinic unit cell featuring perovskite layers formed out of NiO$_6$ octahedra and La layers, separated by rock salt layers.
(b) In-plane and (c) out-of-plane electrical resistivities. The density wave transition temperature $T_\text{DW}$ is highlighted by a vertical dashed line. The inset to (c) shows the temperature dependence of the anisotropy ratio $\gamma$.}
\label{fig:ambient}
\end{figure}

\begin{figure*}[tb]
\centering
\includegraphics[width=175mm]{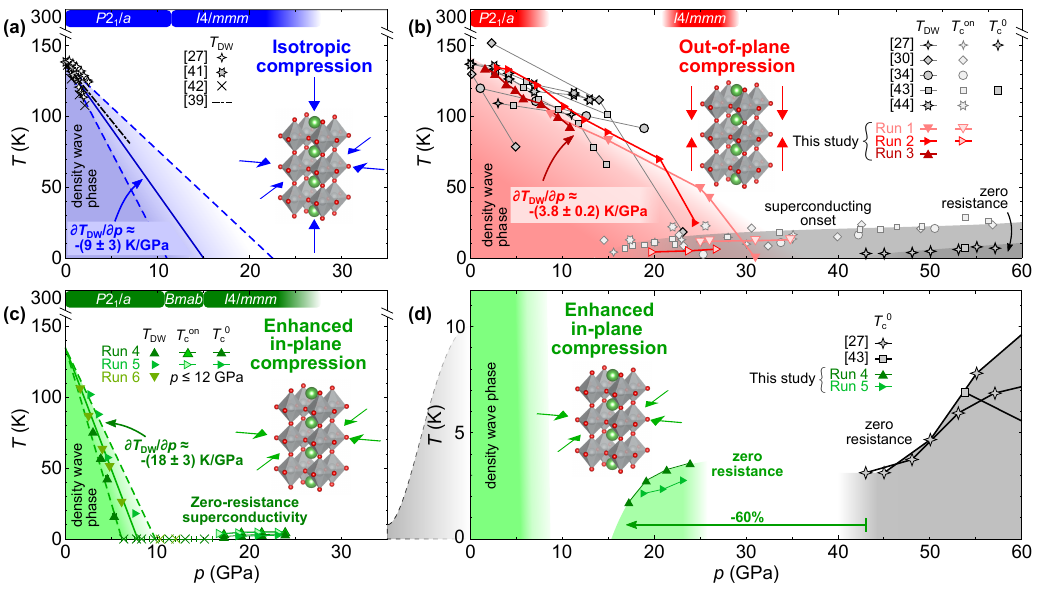}
\caption{Density wave formation and superconductivity in La$_4$Ni$_3$O$_{10}$ under different pressure environments. 
(a) Under (pseudo-) hydrostatic conditions, the density wave phase is suppressed at a rate of approximately $-9$\,K/GPa.
(b) Under dominant out-of-plane compression, the density wave phase is suppressed more slowly, which appears highly consistent with the existing literature. Incomplete superconductivity is observed above $\approx 15-20$\,GPa, and zero-resistance is observed above 40\,GPa only. 
(c) In contrast, under enhanced in-plane compression, the density wave phase is suppressed much faster at a rate of approximately $-18$\,K/GPa, and we observe a zero-resistance state above $\approx 17$\,GPa already.
(d) Zooming into the low-temperature part of panel (c), the zero-resistance transition temperatures $T_c^0$ from our experiments and from existing reports reveal a $60$\,\% lower pressure scale under enhanced in-plane compression.}
\label{fig:phasediag}
\end{figure*}

\section*{Ambient pressures}

Figure~\ref{fig:ambient}(a) shows the result of our scanning transmission electron microscopy experiments which clearly resolve the perovskite blocks formed out of four layers of La atoms with three layers of NiO$_6$ octahedra sandwiched in-between. These blocks are stacked vertically and are separated by a horizontal, in-plane shift, defining the rock salt layer at the interface. Our results reveal an outstanding crystallinity with long-range ordering of La$_4$Ni$_3$O$_{10}$ trilayers only, without any evidence of stacking disorder or intergrowth of other Ruddlesden-Popper phases \cite{Wang2023,Puphal2023,Chen2024a,Li2024b}. 

The high quality of our single crystals is further supported by our electronic transport measurements. Figures~\ref{fig:ambient}(b-c) show the temperature dependence of the resistivity of La$_4$Ni$_3$O$_{10}$, separating the in-plane component $\rho_{ab}$, measured parallel to the NiO$_2$ planes, and the out-of-plane component $\rho_{zz}$, measured perpendicular to the planes. Upon cooling, $\rho_{ab}$ shows continuous metallic resistivity, except for a sharp, peak-like signature at $T_\text{DW} = 135$\,K, which we assign to the known, concomitant spin- and charge-density wave transition \cite{Zhang2020,Li2025a}. Between room-temperature and the lowest temperatures measured, we obtain a residual resistivity ratio RRR $= 12.6$, exceeding any previous report on La$_4$Ni$_3$O$_{10}$. This serves as quantifiable indicator for the high quality of our crystals.

The out-of-plane resistivity $\rho_{zz}$ shows semi-metallic behaviour at room temperature, and increases roughly by a factor four upon cooling into the density wave phase. We determine the resistivity anisotropy $\gamma = \rho_{zz}/\rho_{ab}$, shown as an inset to Fig.~\ref{fig:ambient}(c), which varies between $10^2$ at room temperature and $10^4$ at the lowest temperatures, similar to a recent indirect measurement \cite{Liu2026}. Such large values of $\gamma$ point at a predominantly 2D nature of the electronic structure. This is naturally explained by the quasi-2D lattice structure of La$_4$Ni$_3$O$_{10}$, where the conductive perovskite layers are separated by the insulating rock salt layers, with little effective orbital overlap.

\section*{High pressure behaviour}

Focusing on the density wave transition, previous x-ray diffraction (XRD) and thermal expansion studies linked its onset to a shrinking of the monoclinic unit cell along the $c$ direction, and to an elongation along the $b$ direction \cite{Rout2020,Zhang2020a,Kumar2020,Liu2026}. For the purpose of this report, we will leverage this anisotropic response of the lattice dimensions as an \emph{in-situ} sensor for (non-)hydrostatic pressures. To make this point explicit, we infer from the previous studies that a hypothetically pure uniaxial, out-of-plane compression of the unit cell will support the density wave phase formation and therefore will lead to an increasing transition temperature with pressure \cite{Zhang2020a,Liu2026}. In contrast, assuming a hypothetically pure biaxial, in-plane compression will counteract the lattice changes associated with the onset of the density wave phase, and therefore suppress it with increasing pressure. Naturally, we expect that an ideally hydrostatic environment will be located between these two extreme cases. For this purpose, we summarise in the isotropic phase diagram Fig.~\ref{fig:phasediag}(a) the literature results of thermal expansion measurements on powder samples, and of isotropic, low-pressure $p \lesssim 2$\,GPa transport studies using piston-cylinder cells and liquid pressure media \cite{Wu2001,Zhu2023,Khasanov2026,Rout2020}. They reveal an average suppression rate of the density wave phase of $\partial T_\text{DW}/\partial p \approx -(9 \pm 3)$ K/GPa, which we will take as the hydrostatic reference in the following.

Achieving isotropic pressures beyond 15\,GPa in diamond anvil cells while preserving electrical contacts on the sample is challenging \cite{Zhang2024}. As described in the Methods section and in the Supplemental Material (SM), here we employ diamond anvil cells to generate explicitly non-hydrostatic pressure environments. These pressures generally are not purely uniaxial or biaxial, instead, we generate either dominantly out-of-plane pressures, or enhanced in-plane pressures relative to the sample orientation inside the pressure chamber. Figure~\ref{fig:phasediag}(b)-(d) show the phase diagrams obtained from our experiments, compared to the existing literature \cite{Zhu2023,Zhang2025a,Shi2025,Li2025,Xu2025}. Obviously, the phase diagram alters strongly depending on the compression type, which we will discuss throughout the remainder of this report.

\begin{figure*}[htb]
\centering
\includegraphics[width=160mm]{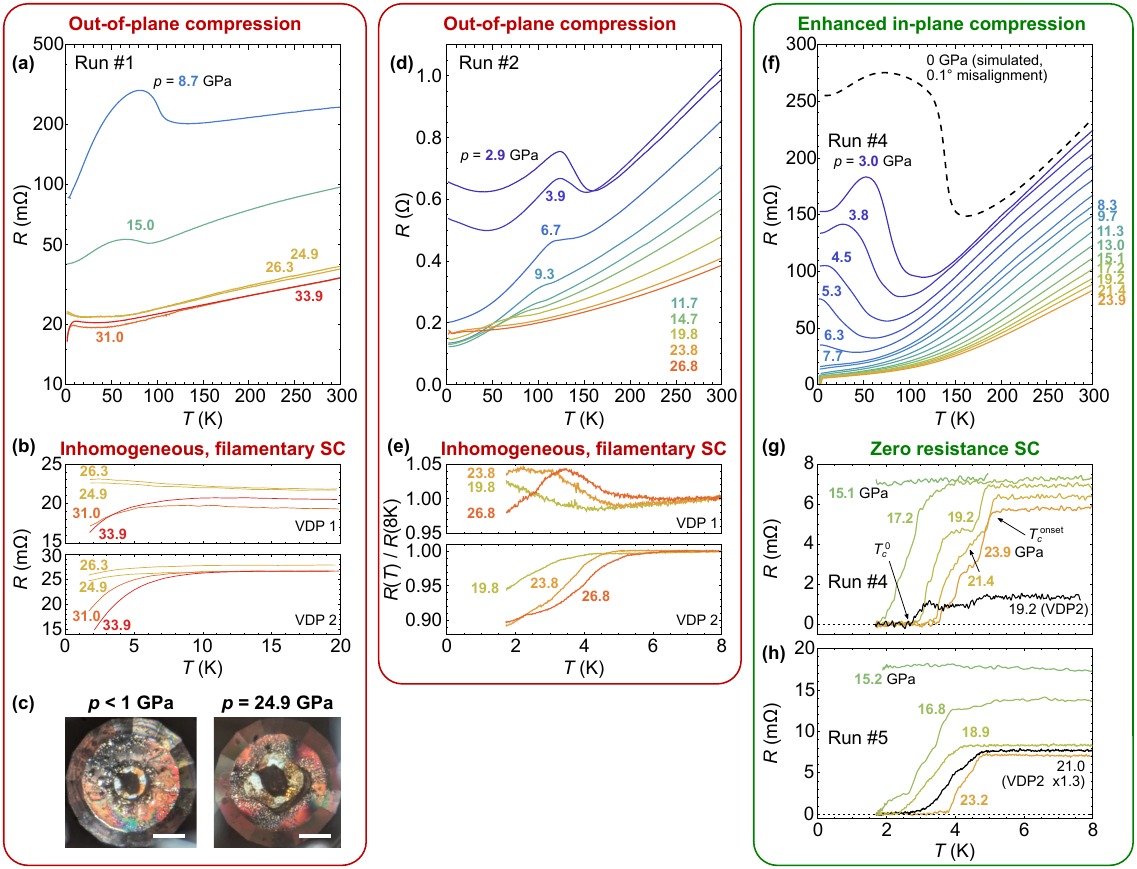}
\caption{High-pressure transport properties of La$_4$Ni$_3$O$_{10}$ under various pressure environments. 
(a,d) Temperature and pressure dependent resistances under dominant out-of-plane compression for runs \#1 and \#2.
(b,e) Zoom into the low-temperature resistance downturns signalling the onset of superconductivity. 
(c) Sample picture showing the strong expansion of the gasket hole and sample under pressure of run \#1, corresponding to a substantial out-of-plane compression.
(f) Temperature and pressure dependent resistance under enhanced in-plane compression.
(g,h) Zoom into the low-temperature resistance featuring a reproducible zero-resistance transition between runs \#4 and \#5.
}
\label{fig:data}
\end{figure*}

\subsection*{Dominant out-of-plane pressure}

Figure~\ref{fig:data} shows the results of our high-pressure experiments. Beginning with the dominant out-of-plane compression in run \#1, we prepared a diamond anvil cell with a gasket in the unstable `thick' limit \cite{Dunstan1989a}. As can be inferred from the optical photographs shown in Fig.~\ref{fig:data}(c), the gasket hole as well as the lateral sample dimensions are found to increase dramatically in-plane with increasing pressure which confirms the expected out-of-plane compression under these conditions. The resistance curves for run \#1 and similarly also for run \#2, shown in Figs.~\ref{fig:data}(a) and (d), respectively, and for run \#3 (shown in the SM) show metallic behaviour at room temperatures for all pressure studied, consistent with the temperature dependence of the ambient pressure $\rho_{ab}$, as shown in Fig.~\ref{fig:ambient}(b). For temperatures below $T \approx 150$\,K, we further identify the signature of the density wave formation, best visible at low pressures only. However, again comparing with the ambient pressure data, it is apparent that this signature is much stronger than in $\rho_{ab}$ at ambient pressure, which suggests that there is a small contribution from $\rho_{zz}$. In the SM, we estimate its relative contribution by taking into account the room-temperature change of the resistance $\partial R/\partial T$, and the ratio between the room temperature resistance and the peak resistance below the density wave formation, which gives a contribution from $\rho_{zz}$ of less than 1\% only, corresponding to a sample misalignment of significantly less than $1^\circ$.

Upon increasing the applied pressure, Figs.~\ref{fig:data}(a) and (d) show that the density wave transition temperature is suppressed continuously, while its signature also broadens and becomes weaker. Still, it remains clearly visible in the data for pressure up to about 20\,GPa. We summarise this pressure dependence of the density wave transition temperature for runs \#1-\#3 in the phase diagram Fig.~\ref{fig:phasediag}(b). We find that the density wave transition is suppressed at a rate $\partial T_\text{DW}/\partial p \approx -3.8$\,K/GPa which is much slower than the hydrostatic reference of $-9$\,K/GPa \cite{Wu2001,Rout2020,Zhu2023,Khasanov2026}, shown in the isotropic phase diagram Fig.~\ref{fig:phasediag}(a). We emphasise that this finding includes the results from run \#1, where the setup most intuitively produces strong out-of-plane pressures because of the unstable gasket. Therefore, the significantly reduced slope $\partial T_\text{DW}/\partial p$ compared to the hydrostatic reference confirms our expectation that dominant out-of-plane pressures tend to support the density wave formation. Finally, we note that this dominant out-of-plane pressure environment reproduces remarkably well the suppression of the density wave phase in essentially all previous high-pressure transport measurements on La$_4$Ni$_3$O$_{10}$ beyond $p \gtrsim 2$\,GPa \cite{Zhu2023,Shi2025,Li2025,Zhang2025a}. This suggests that previous experiments were conducted under similar pressure environments.

Next, Figs.~\ref{fig:data}(b) and (e) show that below $T \approx 10$\,K, and for pressures higher than about 19\,GPa only, we can identify a weak downturn of the resistances measured, similar to the onset of superconductivity reported before. However, even for pressures up to 34\,GPa, the resistance does not reach zero, and the transition appears highly inhomogeneous, with the different van-der-Pauw (VDP) directions giving incongruent results. This suggests that any manifestation of superconductivity under these conditions is filamentary only. Again, this result is fully consistent with previous reports which found superconducting onsets in the same pressure and temperature range, but no evidence for zero resistance below $\approx 42$\,GPa, as shown in Fig.~\ref{fig:phasediag}(b).

\begin{figure*}[tb]
\centering
\includegraphics[width=175mm]{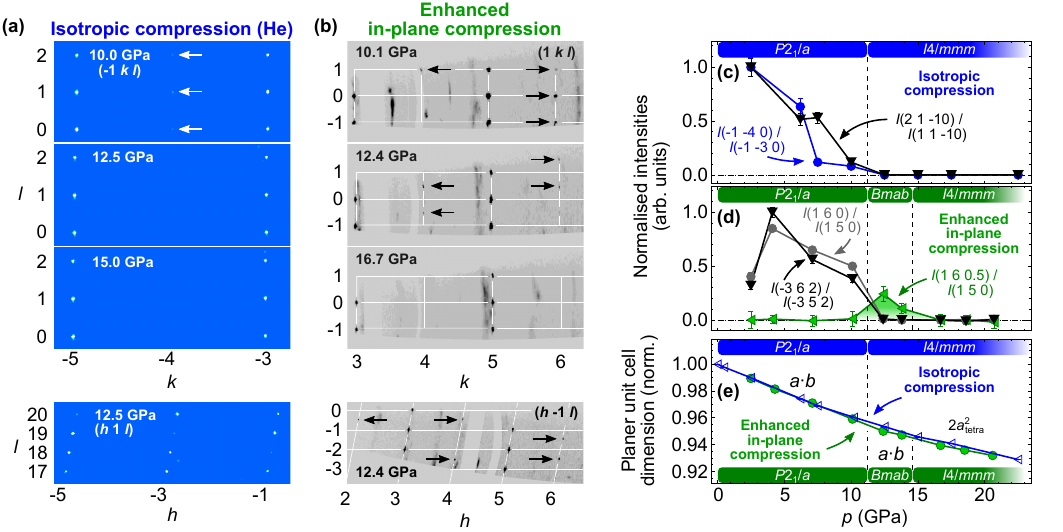}
\caption{Structural properties under pressure. 
(a,b) Reconstructed reciprocal space maps measured by XRD under various pressures set up either (a) for isotropic compression using helium as PTM or a (b) for enhanced in-plane compression. Horizontal arrows mark reflections whose (dis)appearance define structural phase boundaries. 
(c, d) Normalised integrated peak intensities. 
(e) Pressure dependence of the cross section of the unit cell within the $ab$ plane, normalised for either pressure environment against their low-pressure values.
In panels (c-e), vertical dashed lines indicate structural phase transitions between the indicated space groups.
In all panels, the indexation corresponds to the low-pressure monoclinic $P2_1/a$ space group with $a \lesssim b \approx 5.4$\,\AA, $c \approx 14$\,\AA\ and $\beta \approx 101^\circ$ ($Z = 2$).
}
\label{fig:xrd}
\end{figure*}

\subsection*{Enhanced in-plane pressure}

Motivated by the observation of ambient-pressure superconductivity in the thin-film samples of La$_3$Ni$_2$O$_{7}$ under compressive in-plane strain \cite{Ko2025,Zhang2026}, we now discuss the effect of enhancing the in-plane pressures on our La$_4$Ni$_3$O$_{10}$ single crystals, as discussed in the Methods section and in the SM. Figure~\ref{fig:data}(f) shows the evolution of the resistance as measured during run \#4 (we show the results from runs \#5 and \#6 in the SM). Again, we observe metallic behaviour at room-temperatures at all pressures, and we observe the familiar signature of the density wave phase formation. Using the same method as for runs \#1-3, we estimate the sample misalignment to be around $0.1^\circ$. Similarly, as shown in Fig.~\ref{fig:data}(f) as a black dashed line, a simulated ambient-pressure resistance curve assuming a $0.1^\circ$ misalignment reproduces the overall shape of the high-pressure curves remarkably well (see also the SM). Upon increasing the pressure, the density wave is suppressed and its signature disappears completely in the pressure range 8-10\,GPa. This suppression of the density wave phase is manifestly different from the dominant out-of-plane compression cases, where the density wave transition temperature was clearly visible up to 20\,GPa, as evident from Figs.~\ref{fig:data}(a) and (d).

The much faster suppression of the density wave phase is even more apparent from the phase diagram shown in Fig.~\ref{fig:phasediag}(c) which shows that the suppression rate $\partial T_\text{DW}/\partial p$ obtained from run \#4 is as high as $-22$\,K/GPa, far exceeding the hydrostatic reference of $-9$\,K/GPa. Averaging over runs \#4-6, we obtained $\partial T_\text{DW}/\partial p \approx -(18 \pm 3)$\,K/GPa, which is twice as fast as the hydrostatic reference. This result strongly suggests that we succeeded in applying in-plane compressive pressures to our single crystals of La$_4$Ni$_3$O$_{10}$.

Next, Figs.~\ref{fig:data}(g) and (h), and summarised in Figs.~\ref{fig:phasediag}(c) and (d), show that above $p \approx 17$\,GPa, we observe a strong and sharp signature of superconductivity, and the immediate realisation of a zero-resistance state. Moreover, comparing the different VDP current directions, we find that zero resistance is achieved under all directions, further suggesting a homogeneous formation of the superconducting phase. Focusing only on the zero resistance temperature $T_c^0$, Fig.~\ref{fig:phasediag}(d) compares our results to previous reports. The lowest pressures for which we could find $R = 0$ are 17.2\,GPa (run \#4) and 18.9\,GPa (run \#5), which correspond to about 60\% less pressure required compared to previous reports \cite{Zhu2023,Li2025}.

\subsection*{High pressure structures}

Figure~\ref{fig:xrd} compares the results of a series of synchrotron-based high-pressure x-ray diffraction experiments using isotropic compression or enhanced in-plane pressure environments. Here, we present experiments at room temperature to focus solely on the crystalline response in the absence of the low-temperature density wave phase. Further data is discussed in the SM.

At ambient pressures, we confirm the previously reported monoclinic $P2_1/a$ space group with $a = 5.4047(4)$\,\AA, $b = 5.45129(18)$\,\AA, $c = 14.1685(3)$\,\AA\ and $\beta = 100.853(4)^\circ$ ($Z = 2$). We observe a number of weak reflections at half-integer $l$ if $h + k$ is odd, which can be accounted for through pseudo-merohedral twinning. We therefore do not interpret them as signatures for a doubling of the unit cell dimension along the $c$ direction, which otherwise might point at orthorhombic or weakly monoclinic space groups \cite{Zhang2020a,Yuan2024}. In contrast, under the highest pressures studied beyond 20\,GPa, we resolve the tetragonal $I4/mmm$ space group irrespective of the pressure environment. This space group was also found for the sample used for run \#4 described above, as shown in the SM. 

Across the monoclinic-to-tetragonal structural transition at intermediate pressures, we identify different behaviours depending on the pressure environment. For the isotropic case, the transition from $P2_1/a$ to $I4/mmm$ is indicated by the disappearance of several reflections, as shown in Fig.~\ref{fig:xrd}(a) and (c), consistent with the increasing lattice symmetry. This allows us to locate the monoclinic-to-tetragonal transition around $11$\,GPa. For enhanced in-plane pressures, we find that the suppression of the monoclinic phase occurs essentially at the same pressure as for the isotropic phase. However, we find that the formation of the tetragonal phase is pre-empted by an intermediate orthorhombic structure. Figure~\ref{fig:xrd}(b) and (d) show that beyond 10\,GPa, the monoclinic $(1\ 4\ l)$ and $(1\ 6\ l)$ reflections disappear for integer $l$, but new reflections appear for half-integer $l$, which in turn disappear again once the tetragonal phase is reached beyond 16\,GPa. We can rule out that these half-integer reflections are the result of pseudo-merohedral twinning because the corresponding monoclinic twinning partners are absent, and because these half-integer reflections do not occur at any lower pressure. Instead, the new reflections are consistent with the orthorhombic $Bmab$ structure with a doubling of the unit cell along the $c$ direction, which was previously reported in the same pressure range \cite{Li2025} and observed for certain crystals at ambient conditions and room temperature \cite{Zhang2020a,Yuan2024}. This observation shows that the monoclinic-to-tetragonal structural transition is easily perturbed by the pressure environment and that the orthorhombic phase is energetically in close proximity \cite{Ramakrishnan2026}.

To conclude this section, we provide a rough estimate for the pressure anisotropy under the enhanced in-plane compression. Figure~\ref{fig:xrd}(e) shows how the planar cross-section of the unit cell shrinks under pressure, as obtained from the lattice parameters $a$ and $b$ (see also the SM). We read off that the enhanced in-plane pressure environment leads to an additional shrinking of the unit cell within the high-pressure tetragonal phase, by approximately $0.2-0.3$ percentage points compared to the isotropic environment. Note that the pressures applied here are much smaller than the bulk modulus of La$_4$Ni$_3$O$_{10}$, which we obtain as $B(p) \approx 187\,\text{GPa} + 3.6 p$ in the high-pressure phase (shown in the SM), similar to a previous report \cite{Li2024a}. Therefore, we assume a linear relation between the directional dependent length changes $\delta l_i/l_i$ and the corresponding non-hydrostatic pressure components $p_i$ for $i \in \{x, y, z\}$, from which we estimate the diagonal components of the pressure tensor $p_{ii} \approx p_0 \cdot (1.04(1), 1.04(1), 0.92(2))$. These values corresponds to a pressure anisotropy of approximately $12$\,\% between the in-plane and out-of-plane pressure components, valid for a hydrostatic $p_0$ in the range $15-20$\,GPa.

\section*{Discussion}

Our experiments comparing the effect of out-of-plane versus in-plane pressures on single crystals of La$_4$Ni$_3$O$_{10}$ lead to a number of profound discoveries beyond the remarkable realisation of a reproducible zero-resistive state discussed above:

Firstly, under enhanced in-plane compression, we observe an intermediate pressure regime between $p \approx 10$\,GPa and $p \approx 17$\,GPa for which the density wave phase has been suppressed already, but where we do not observe any signature of superconductivity. In contrast, above 17\,GPa, we find a zero-resistance state at low temperatures, which develops a sharp superconducting transition width $\Delta T_c \lesssim 1$\,K above 20\,GPa. Moreover, the reproducible separation of superconductivity and the density wave phase clearly show that superconductivity is not driven by soft, quantum critical fluctuations of the density wave phase order parameter, for example emerging from a putative quantum critical point \cite{Shi2025,Inoue2026}. However, alternative scenarios in which non-critical fluctuations away from the quantum phase transition are assisting with the pairing, or scenarios where the density wave phase and superconductivity are competitors, should still be considered. Moreover, we cannot rule out the persistence of quasi-2D short-range density wave order which might not generate a sharp imprint in the resistivity in this pressure regime \cite{Matsuoka2018,Lee2021,Keimer2015}.

Secondly, our comparative transport and XRD experiments under consistent pressure environments support the prevailing understanding that superconductivity in the Ruddlesden-Popper nickelates does not occur in the low-pressure monoclinic phase with strong octahedral buckling. We find that this phase is suppressed at a pressure $p \approx 10-11$\,GPa at room temperature, irrespective of the pressure environment. Even though the transition pressure could increase by $3-4$\,GPa towards lower temperatures \cite{Ramakrishnan2026}, it remains much smaller compared to the onset of superconductivity around 17\,GPa. Our results leave open the possibility that superconductivity may occur in an intermediate orthorhombic phase, e.g. with $Bmab$ symmetry, for which the octahedral buckling is reduced but remains finite \cite{Li2025}.

Thirdly, our results strongly suggest that previous high-pressure transport measurements experienced strong but unreported uniaxial pressure components. This challenges the interpretation of previous comparative transport and structural studies \cite{Sun2023,Zhu2023,Zhang2025,Zhang2026}. Similarly, the previously reported coexistence of the density wave phase, as detected by transport measurements on one hand, and bulk superconductivity, as detected by contactless susceptibility measurements on the other, needs to be critically assessed \cite{Zhang2025}. For example, it appears very unusual that a large diamagnetic signal was observed at lower pressure and/or higher temperatures than zero-resistance superconductivity \cite{Zhu2023,Zhang2025}. We comment in the SM about a potential origin of such observations.

To conclude, our discoveries reported here have profound implications on the field of superconducting nickelates and beyond. Firstly, our observation of a zero-resistance state in La$_4$Ni$_3$O$_{10}$ at pressures as low as $p \approx 17$\,GPa solves the puzzles of seemingly different pressure scales required for superconductivity in bilayer La$_3$Ni$_2$O$_7$ and trilayer La$_4$Ni$_3$O$_{10}$. Instead, we show that both systems become superconducting under the same pressure scale, thereby providing a route towards a unified description of superconductivity in both systems \cite{Jiang2025}. Secondly, by comparing the effects of in-plane and out-of-plane pressures, our results provide central benchmarks for theoretical studies, weighing the relative importance of orbital hybridisation and/or superexchange between planar Ni $d_{x^2-y^2}$ orbitals on one side, and apical Ni $d_{z^2}$ orbitals on the other \cite{Samanta2025,Jiang2026}. Thirdly, our results show that boosting the biaxial, in-plane pressure is a simple, yet powerful \emph{in-situ} method for enhancing superconductivity in the Ruddlesden-Popper nickelates. Our results draw a connection to the bilayer systems (La,Pr)$_3$Ni$_2$O$_7$ and (La,Sm)$_3$Ni$_2$O$_7$ systems, where the cation substitution leads to a reduced in-plane $a$ and $b$ parameters, and where signatures of bulk superconductivity were observed under pressure, despite a strong broadening of the La nuclear quadrupole resonance (NQR) line \cite{Wang2024,Li2026}. Therefore, similar studies on related RE$_3$Ni$_2$O$_7$ samples are warranted to further study this effect. In the same spirit, our results connect bulk crystals and thin-film samples, which become superconducting under compressive biaxial strain \cite{Zhang2026}. Fourthly, our results provide a natural connection to several families of the cuprate superconductors, where biaxial pressures were found to enhance $T_c$ even in optimally doped samples, whereas out-of-plane pressures were found to decrease $T_c$ \cite{Hardy2010}. This similarity in the two types of unconventional superconducting oxide families could point at a common mechanism behind the superconducting pairing, and might be the starting point for future comparative studies.

\section*{Methods}

\subsection*{Synthesis}

For the synthesis of La$_4$Ni$_3$O$_{10-x}$ single crystals, we utilize the optical floating zone (OFZ) technique under high oxygen pressure. We nominally adopt the same growth conditions as reported in refs. \cite{Zhang2020,Zhu2023}, including an oxygen partial pressure of 20 bar. Further details are provided in the SM.

\subsection*{Ambient and high pressure electronic transport measurements}

Pressures exceeding 30 GPa were generated using home-made Merrill-Bassett type diamond anvil cell with culet diameters ranging between 300\,$\mu$m and 500\,$\mu$m, with a single bevel. We employed two different kinds of gaskets: (1) Rhenium gaskets were pre-indented to approx 25-35\,$\mu$m depending on the culet size, subsequently insulated using a mixture of Stycast 1266 and Al$_2$O$_3$ powder, and the final pressure chamber was drilled to a diameter between 100-150$\mu$m. Since Rhenium is well known to become superconducting at low temperatures, especially under high non-hydrostatic pressures, we checked after every run that no electrical contact occurred between the gasket and the electrodes leading to the sample. (2) A composite gasket was produced by pre-indenting Rhenium or tungsten gaskets in the same way as in (1), but subsequently, the gasket was drilled with a drill bit exceeding the culet diameter. The hole was filled with a similar mixture of Stycast 1266 and Al$_2$O$_3$ powder, which was pressed between the diamonds, cured, and subsequently drilled to produce the sample chamber. 

We use the standard ruby fluorescence method for the pressure determination at room temperature, which revealed sharp and well separated R1 and R2 lines up to the highest pressures achieved. Typically, we placed 3-4 rubies for each experiment in the pressure chamber, distributed such that one ruby would almost touch the sample, one would almost touch the gasket, and the others were placed randomly between the sample and the gasket. This setup allows us to monitor the local pressure distribution in the pressure chamber. The values reported here are the ones determined from the ruby closest to the sample.

For the transport measurements, samples were cleaved carefully to expose a shiny, terrace-free surface, which we reliably found to be the $ab$ plane. For the ambient pressure measurements, electrical contacts were produced using Ag epoxy \cite{Reiss2026}, and we used a standard longitudinal five-contact setup for measuring the in-plane sample resistance. The out-of-plane resistivity was measured using a van-der-Pauw geometry on a cuboid sample. For the high pressure measurements, we created electrodes in a van-der-Pauw geometry inside the pressure chamber. For all measurements, we employed either the PPMS resistance option, or the LockIn method combined with a Stanford Research SR554 transformer. Excitations currents were typically 100\,$\mu$A below 10\,K, and up to 1\,mA towards room temperature. 

For the ambient pressure measurements and for runs \#2-\#6, non-annealed crystal were used from the same boule in order to minimise any potential sample-to-sample variation. For comparison, the sample for run \#1 was annealed in oxygen (see the SM for further details), but we see no significant changes in transport properties or their pressure dependence, except for a slightly higher superconducting onset temperature under pressure. The sample from run \#4 was investigated using XRD after the last transport measurement, whilst still located in the same pressure cell (see the SM). A small piece originally broken off from the sample used in run \#5 was used for the pressure dependent XRD study.

\subsection*{Electron microscopy}
La$_4$Ni$_3$O$_{10-x}$ single crystals were investigated by scanning transmission electron microscopy (STEM). Electron-transparent specimens were prepared on a Thermo Fisher Scios I FIB using the standard liftout method. The lateral dimensions of the specimens were 20 $\mu$m by 1.5 $\mu$m with thicknesses between 50 and 100 nm. The STEM images were recorded by a probe aberration-corrected JEOL JEM-ARM200F scanning transmission electron microscope equipped with a cold-field emission electron source and a probe Cs corrector (DCOR, CEOS GmbH) was used at 200 kV. STEM imaging was performed at probe semiconvergence angles of 20 mrad, resulting in probe sizes of 0.8 {\AA}. Inner and outer collection angles for STEM-HAADF were 75 and 310 mrad, respectively. To improve the signal-to-noise ratio of the data while minimizing sample damage, a high-speed time series was recorded (2 $\mu$s per pixel) and was then aligned and summed. 

\subsection*{XRD}
The single crystal XRD data under pressure were collected at the beamline ID27 at the European Synchrotron Radiation Facility (ESRF). For the isotropic pressure generation, we used membrane driven diamond anvil cells (DAC) and helium as the pressure transmitting medium. Natural single crystalline diamonds with Boehler-Almax design and culets of 500 $\mu$m in size were mounted on seats with a modified Boehler-Almax design and an opening of 70$^\circ$. Stainless steel was used as the gasket material. 200 $\mu$m thick plates were pre-indented to 100 $\mu$m. The sample chamber was created by laser drilling a hole of 300 $\mu$m in diameter. The cell was loaded with compressed He and closed at 1400 bar. The applied pressure was determined from ruby fluorescence. 

For the enhanced in-plane pressures, we used the same pressure cells as for the transport measurements, with 500 $\mu$m culets and an opening angle of 30$^\circ$. A Rhenium gasket was pre-indented to 45 $\mu$m thickness and a 150\,$\mu$m hole was drilled as the sample chamber. We used a small sample that was cleaved off from the same crystal as for run \#5 (see Fig.~\ref{fig:data}). We also measured the same sample as in run \#4 still located in the pressure cell used for the transport measurements (shown in the SM).

Data was collected in continuous $\phi$ rotation over 68$^\circ$ (or 30$^\circ$) with 0.25$^\circ$ (0.5$^\circ$) steps for the isotropic (enhanced in-plane) compression case, respectively, performing a shutterless readout of the EIGER2 X CdTe 9M detector (Dectris, Switzerland). The detector distance was calibrated using CeO$_2$ powder to 185.5 mm (isotropic case) and 298 mm (enhanced in-plane), and the diffraction geometry calibrated using a natural vanadinite single crystal. The flat-field of the detector was calibrated using air scattering at the employed x-ray energies.
The key experimental results were obtained from La$_4$Ni$_3$O$_{10}$ crystals with sizes of approximately $50 \times 30 \times 20$ $\mu$m$^3$, using monochromatic x-rays with a wavelength of $0.22290$ {\AA}, collecting at least 4000 (isotropic) or a few hundred (enhanced in-plane) distinct indexed Bragg reflexes at each applied pressure.

Our refined lattice parameters are in the SM in comparison to literature results, revealing a general transition around 10-15 GPa. A certain spread depending on oxygen content and isotropy of pressure can be expected. Overall, our result lies within the range of all reports.

\section*{Acknowledgements}

We acknowledge the ESRF within the in-house proposals IH-HC-3746 and IH-HC-4342 for provision of synchrotron radiation. We thank Jeroen Jacobs for preparation and gas loading of the high-pressure cells. We also thank Markus Dueller for technical assistance during the development of the pressure cells in Stuttgart, and Marion Hagel and Thomas Reindl from the Nanostructuring Lab in Stuttgart for producing Au electrodes on the diamond anvils.

\section*{Author contributions statement}
High pressure transport measurements were conceived and carried out by P.R.. Thin polished or cleaved flakes were prepared by P.R. and A.S., and oriented by P.P. and J.N.. Crystals were grown and annealed by P.P.. Single-crystal XRD under pressure was done at ESRF by B.W., P.R., V.S., T.P, and M.K.. P.P. and P.R. analyzed the data. Standard characterization was done by P.P.. STEM experiments were done by P.S.L. and Y.-M.W. under the supervision of P.A.A. and Y.E.S..  P.R., and P.P. wrote the manuscript. All authors reviewed the manuscript. 

\section*{Data availability}
The high-pressure x-ray diffraction data from ID27 is available xxx.

\section*{Competing interests}
The authors declare no competing interests.

\section*{Additional information}
The online version contains supplemental material available at xxx.

\bibliography{bib}

@article{Jiang2026,
archivePrefix = {arXiv},
arxivId = {2603.13808},
author = {Jiang, Guiwen and Si, Liang and Sawatzky, George A. and Jiang, Mi},
eprint = {2603.13808},
journal = {arXiv (online)},
month = {jun},
title = {{Pressure induced redistribution of oxygen hole states in La$_{4}$Ni$_{3}$O$_{10}$}},
url = {http://arxiv.org/abs/2603.13808},
year = {2026}
}

@article{Dunstan1989a,
author = {Dunstan, D. J.},
doi = {10.1063/1.1140442},
issn = {0034-6748},
journal = {Review of Scientific Instruments},
month = {dec},
number = {12},
pages = {3789--3795},
title = {{Theory of the gasket in diamond anvil high‐pressure cells}},
url = {http://aip.scitation.org/doi/10.1063/1.1140442},
volume = {60},
year = {1989}
}

@article{Li2025a,
author = {Li, Mingzhe and Gong, Jiashuo and Zhu, Yinghao and Chen, Ziyuan and Zhang, Jiakang and Zhang, Enkang and Li, Yuanji and Yin, Ruotong and Wang, Shiyuan and Zhao, Jun and Feng, Dong-Lai and Du, Zengyi and Yan, Ya-Jun},
doi = {10.1103/2p56-xl41},
issn = {2469-9950},
journal = {Physical Review B},
month = {jul},
number = {4},
pages = {045132},
title = {{Direct visualization of an incommensurate unidirectional charge density wave in La$_4$Ni$_3$O$_{10}$}},
volume = {112},
year = {2025}
}

@article{Ko2025,
author = {Ko, Eun Kyo and Yu, Yijun and Liu, Yidi and Bhatt, Lopa and Li, Jiarui and Thampy, Vivek and Kuo, Cheng-Tai and Wang, Bai Yang and Lee, Yonghun and Lee, Kyuho and Lee, Jun-Sik and Goodge, Berit H. and Muller, David A. and Hwang, Harold Y.},
doi = {10.1038/s41586-024-08525-3},
isbn = {4158602408},
issn = {0028-0836},
journal = {Nature},
month = {feb},
number = {8052},
pages = {935--940},
pmid = {39701131},
publisher = {Springer US},
title = {{Signatures of ambient pressure superconductivity in thin film La$_3$Ni$_2$O$_7$}},
url = {https://www.nature.com/articles/s41586-024-08525-3},
volume = {638},
year = {2025}
}

@article{Reiss2026,
archivePrefix = {arXiv},
arxivId = {2606.27040},
author = {Reiss, P. and Shevchenko, A. and Lizama, P. S. and Nuss, J. and Dinnebier, R. and van Aken, P. A. and Hepting, M. and Isobe, M. and Suyolcu, Y. E. and Takagi, H. and Keimer, B. and Puphal, P.},
eprint = {2606.27040},
journal = {arXiv (online)},
title = {{High temperature transitions in Ruddlesden-Popper nickelates La$_{n+1}$Ni$_{n}$O$_{3n+1}$}},
url = {http://arxiv.org/abs/2606.27040},
year = {2026}
}

@article{Hardy2010,
author = {Hardy, F. and Hillier, N. J. and Meingast, C. and Colson, D. and Li, Y. and Bari{\v{s}}i{\'{c}}, N. and Yu, G. and Zhao, X. and Greven, M. and Schilling, J. S.},
doi = {10.1103/PhysRevLett.105.167002},
issn = {0031-9007},
journal = {Physical Review Letters},
month = {oct},
number = {16},
pages = {167002},
title = {{Enhancement of the Critical Temperature of HgBa$_2$CuO$_{4+\delta}$ by Applying Uniaxial and Hydrostatic Pressure: Implications for a Universal Trend in Cuprate Superconductors}},
url = {https://link.aps.org/doi/10.1103/PhysRevLett.105.167002},
volume = {105},
year = {2010}
}

@article{Kumar2020,
author = {Kumar, Susmit and Fjellv{\aa}g, {\O}ystein and Sj{\aa}stad, Anja Olafsen and Fjellv{\aa}g, Helmer},
doi = {10.1016/j.jmmm.2019.165915},
issn = {03048853},
journal = {Journal of Magnetism and Magnetic Materials},
month = {feb},
number = {2007},
pages = {165915},
title = {{Physical properties of Ruddlesden-Popper ($n$ = 3) nickelate: La$_4$Ni$_3$O$_{10}$}},
url = {https://linkinghub.elsevier.com/retrieve/pii/S0304885319320396},
volume = {496},
year = {2020}
}

@article{Zhang2026,
author = {Zhang, Yang and Lin, Fang-Ling and Maier, Thomas A and Dagotto, Elbio},
doi = {10.1088/1361-6633/ae9aae},
eprint = {2604.18385},
journal = {Reports on Progress in Physics},
month = {apr},
number = {9},
pages = {096501},
title = {{Superconductivity in Ruddlesden-Popper nickelates: a review of recent progress, focusing on thin films}},
url = {https://doi.org/10.1088/1361-6633/ae9aae},
volume = {89},
year = {2026}
}

@article{Mandyam2026,
author = {Mandyam, Srinivas V. and Wang, E. and Wang, Zhipan and Chen, Bijuan and Jayarama, Nishan C. and Gupta, Anmay and Riesel, Eric A. and Levitas, Valery I. and Laumann, Christopher R. and Yao, Norman Y.},
doi = {10.1038/s41586-025-10095-x},
issn = {0028-0836},
journal = {Nature},
month = {mar},
number = {8104},
pages = {54--60},
publisher = {Springer US},
title = {{Uncovering origins of heterogeneous superconductivity in La$_3$Ni$_2$O$_7$}},
url = {https://www.nature.com/articles/s41586-025-10095-x},
volume = {651},
year = {2026}
}

@article{Zhang2020a,
author = {Zhang, Junjie and Zheng, Hong and Chen, Yu-Sheng and Ren, Yang and Yonemura, Masao and Huq, Ashfia and Mitchell, J. F.},
doi = {10.1103/PhysRevMaterials.4.083402},
issn = {2475-9953},
journal = {Physical Review Materials},
month = {aug},
number = {8},
pages = {083402},
title = {{High oxygen pressure floating zone growth and crystal structure of the metallic nickelates R$_4$Ni$_3$O$_{10}$ (R = La, Pr)}},
url = {http://arxiv.org/abs/1904.10048},
volume = {4},
year = {2020}
}

@article{Liu2026,
archivePrefix = {arXiv},
arxivId = {2605.18524},
author = {Liu, Feiyang and Chen, Lixing and Zhang, Enkang and Zhang, Ying-jie and Zhao, Jun},
eprint = {2605.18524},
journal = {arXiv (online)},
month = {may},
title = {{Interlayer electronic coherence links magnetism and superconductivity in Ruddlesden-Popper nickelates}},
url = {http://arxiv.org/abs/2605.18524},
year = {2026}
}

@article{Rout2020,
author = {Rout, Dibyata and Mudi, Sanchayeta Ranajit and Hoffmann, Marco and Spachmann, Sven and Klingeler, R{\"{u}}diger and Singh, Surjeet},
doi = {10.1103/PhysRevB.102.195144},
issn = {2469-9950},
journal = {Physical Review B},
month = {nov},
number = {19},
pages = {195144},
publisher = {American Physical Society},
title = {{Structural and physical properties of trilayer nickelates $R_4$Ni$_3$O$_{10}$ ($R =$ La, Pr, and Nd)}},
url = {https://link.aps.org/doi/10.1103/PhysRevB.102.195144},
volume = {102},
year = {2020}
}

@article{Zhu2023,
author = {Zhu, Yinghao and Peng, Di and Zhang, Enkang and Pan, Bingying and Chen, Xu and Chen, Lixing and Ren, Huifen and Liu, Feiyang and Hao, Yiqing and Li, Nana and Xing, Zhenfang and Lan, Fujun and Han, Jiyuan and Wang, Junjie and Jia, Donghan and Wo, Hongliang and Gu, Yiqing and Gu, Yimeng and Ji, Li and Wang, Wenbin and Gou, Huiyang and Shen, Yao and Ying, Tianping and Chen, Xiaolong and Yang, Wenge and Cao, Huibo and Zheng, Changlin and Zeng, Qiaoshi and Guo, Jian-gang and Zhao, Jun},
doi = {10.1038/s41586-024-07553-3},
issn = {0028-0836},
journal = {Nature},
month = {jul},
number = {8021},
pages = {531--536},
title = {{Superconductivity in pressurized trilayer La$_4$Ni$_3$O$_{10-\delta}$ single crystals}},
url = {https://www.nature.com/articles/s41586-024-07553-3},
volume = {631},
year = {2024}
}

@article{Li2024a,
author = {Li, Jingyuan and Chen, Cui-Qun and Huang, Chaoxin and Han, Yifeng and Huo, Mengwu and Huang, Xing and Ma, Peiyue and Qiu, Zhengyang and Chen, Junfeng and Hu, Xunwu and Chen, Lan and Xie, Tao and Shen, Bing and Sun, Hualei and Yao, Dao-Xin and Wang, Meng},
doi = {10.1007/s11433-023-2329-x},
issn = {1674-7348},
journal = {Science China Physics, Mechanics \& Astronomy},
month = {nov},
number = {11},
pages = {117403},
title = {{Structural transition, electric transport, and electronic structures in the compressed trilayer nickelate La$_4$Ni$_3$O$_{10}$}},
url = {https://link.springer.com/10.1007/s11433-023-2329-x},
volume = {67},
year = {2024}
}

@article{Xu2025,
author = {Xu, Shuxiang and Wang, Hao and Huo, Mengwu and Hu, Deyuan and Wu, Qiong and Yue, Li and Wu, Dong and Wang, Meng and Dong, Tao and Wang, Nanlin},
doi = {10.1038/s41467-025-62294-9},
issn = {2041-1723},
journal = {Nature Communications},
month = {jul},
number = {1},
pages = {7039},
publisher = {Springer US},
title = {{Collapse of density wave and emergence of superconductivity in pressurized-La$_4$Ni$_3$O$_{10}$ evidenced by ultrafast spectroscopy}},
url = {https://www.nature.com/articles/s41467-025-62294-9},
volume = {16},
year = {2025}
}

@Article{Ramakrishnan2026,
  author    = {Ramakrishnan, Sitaram and Gao, Yingzheng and Olevano, Valerio and Pachoud, Elise and Hadj-Azzem, Abdellali and Garbarino, Gaston and d’Astuto, Matteo and Perez, Olivier and Pautrat, Alain and Valenti, Diego and Quénot, Matthieu and Pairis, Sébastien and Chernyshov, Dmitry and Noohinejad, Leila and Paulmann, Carsten and Bulled, Johnathan and Bosak, Alexei and van Smaalen, Sander and Toulemonde, Pierre and Méasson, Marie-Aude and Rodière, Pierre},
  journal   = {Communications Materials},
  title     = {{Tracing the horizon of tetragonal-to-monoclinic distortion in pressurized trilayer nickelate La$_4$Ni$_3$O$_{10}$}},
  year      = {2026},
  issn      = {2662-4443},
  month     = {July},
  doi       = {10.1038/s43246-026-01297-5},
  publisher = {Springer Science and Business Media LLC},
}

@article{Wu2001,
author = {Wu, Guoqing and Neumeier, J. J. and Hundley, M. F.},
doi = {10.1103/PhysRevB.63.245120},
issn = {0163-1829},
journal = {Physical Review B},
month = {jun},
number = {24},
pages = {245120},
title = {{Magnetic susceptibility, heat capacity, and pressure dependence of the electrical resistivity of La$_3$Ni$_2$O$_7$ and La$_4$Ni$_3$O$_{10}$}},
url = {https://link.aps.org/doi/10.1103/PhysRevB.63.245120},
volume = {63},
year = {2001}
}

@article{Li2025,
author = {Li, Feiyu and Hao, Yinqiao and Guo, Ning and Jia, Donghan and Dou, Jie and Peng, Di and Zhang, Gongting and Xiao, Lingfeng and Zhang, Junjie Jian and Zhou, Wenju and Huang, Yehua and Wang, Xinyu and Guo, Zhenyuan and Mezouar, Mohamed and {Elias F. S. Rodrigues}, Jo{\~{a}}o and Luo, Jun and Yang, Jie and Zeng, Qiaoshi and Zhou, Rui and Gou, Huiyang and Zheng, Qiang and Liu, Guangtao and Zhang, Junjie Jian},
doi = {https://doi.org/10.1002/adma.202507365},
issn = {0935-9648},
journal = {Advanced Materials},
month = {nov},
number = {44},
pages = {e07365},
title = {{Single-Crystal Structure Determination of Superconducting La$_4$Ni$_3$O$_{10-\delta}$ under High Pressure}},
url = {https://advanced.onlinelibrary.wiley.com/doi/abs/10.1002/adma.202507365},
volume = {37},
year = {2025}
}

@article{Sun2023,
author = {Sun, Hualei and Huo, Mengwu and Hu, Xunwu and Li, Jingyuan and Liu, Zengjia and Han, Yifeng and Tang, Lingyun and Mao, Zhongquan and Yang, Pengtao and Wang, Bosen and Cheng, Jinguang and Yao, Dao-xin and Zhang, Guang-ming and Wang, Meng},
doi = {10.1038/s41586-023-06408-7},
issn = {0028-0836},
journal = {Nature},
month = {sep},
number = {7979},
pages = {493--498},
publisher = {Springer US},
title = {{Signatures of superconductivity near 80 K in a nickelate under high pressure}},
url = {https://www.nature.com/articles/s41586-023-06408-7},
volume = {621},
year = {2023}
}

@article{Zhang2020,
author = {Zhang, Junjie and Phelan, D. and Botana, A. S. and Chen, Yu-Sheng and Zheng, Hong and Krogstad, M. and Wang, Suyin Grass and Qiu, Yiming and Rodriguez-Rivera, J. A. and Osborn, R. and Rosenkranz, S. and Norman, M. R. and Mitchell, J. F.},
doi = {10.1038/s41467-020-19836-0},
issn = {2041-1723},
journal = {Nature Communications},
month = {nov},
number = {1},
pages = {6003},
pmid = {33243978},
publisher = {Springer US},
title = {{Intertwined density waves in a metallic nickelate}},
url = {https://www.nature.com/articles/s41467-020-19836-0},
volume = {11},
year = {2020}
}

@article{Shi2025,
author = {Shi, Mengzhu and Li, Yikang and Wang, Yuxing and Peng, Di and Yang, Shaohua and Li, Houpu and Fan, Kaibao and Jiang, Kun and He, Junfeng and Zeng, Qiaoshi and Song, Dongsheng and Ge, Binghui and Xiang, Ziji and Wang, Zhenyu and Ying, Jianjun and Wu, Tao and Chen, Xianhui},
doi = {10.1038/s41467-025-57264-0},
issn = {2041-1723},
journal = {Nature Communications},
month = {mar},
number = {1},
pages = {2887},
publisher = {Springer US},
title = {{Absence of superconductivity and density-wave transition in ambient-pressure tetragonal La$_4$Ni$_3$O$_{10}$}},
url = {https://www.nature.com/articles/s41467-025-57264-0},
volume = {16},
year = {2025}
}

@article{Jiang2025,
archivePrefix = {arXiv},
arxivId = {2505.13442},
journal = {arXiv (online)},
author = {Jiang, Ruoshi and Fan, Zhiyu and Monserrat, Bartomeu and Ku, Wei},
eprint = {2505.13442},
month = {oct},
title = {{Pressure-induced trans-proximate correlation in La$_4$Ni$_3$O$_{10}$ and possible routes to enhance its superconductivity}},
url = {http://arxiv.org/abs/2505.13442},
year = {2025}
}

@article{Li2026,
author = {Li, Feiyu and Xing, Zhenfang and Peng, Di and Dou, Jie and Guo, Ning and Ma, Liang and Zhang, Yulin and Wang, Lingzhen and Luo, Jun and Yang, Jie and Zhang, Jian Junjie and Chang, Tieyan and Chen, Yu-Sheng and Cai, Weizhao and Cheng, Jinguang and Wang, Yuzhu and Liu, Yuxin and Luo, Tao and Hirao, Naohisa and Matsuoka, Takahiro and Kadobayashi, Hirokazu and Zeng, Zhidan and Zheng, Qiang and Zhou, Rui and Zeng, Qiaoshi and Tao, Xutang and Zhang, Jian Junjie},
doi = {10.1038/s41586-025-09954-4},
issn = {0028-0836},
journal = {Nature},
month = {jan},
number = {8098},
pages = {871--878},
publisher = {Springer US},
title = {{Bulk superconductivity up to 96 K in pressurized nickelate single crystals}},
url = {https://www.nature.com/articles/s41586-025-09954-4},
volume = {649},
year = {2026}
}

@article{Zhang2025,
author = {Zhang, Enkang and Peng, Di and Zhu, Yinghao and Chen, Lixing and Cui, Bingkun and Wang, Xingya and Wang, Wenbin and Zeng, Qiaoshi and Zhao, Jun},
doi = {10.1103/PhysRevX.15.021008},
issn = {2160-3308},
journal = {Physical Review X},
month = {apr},
number = {2},
pages = {021008},
publisher = {American Physical Society},
title = {{Bulk Superconductivity in Pressurized Trilayer Nickelate Pr$_4$Ni$_3$O$_{10}$ Single Crystals}},
url = {https://link.aps.org/doi/10.1103/PhysRevX.15.021008},
volume = {15},
year = {2025}
}

@article{Huang2025,
archivePrefix = {arXiv},
arxivId = {2510.12250},
journal = {arXiv (online)},
author = {Huang, Chaoxin and Li, Jingyuan and Huang, Xing and Zhang, Hengyuan and Hu, Deyuan and Huo, Mengwu and Chen, Xiang and Chen, Zhen and Sun, Hualei and Wang, Meng},
eprint = {2510.12250},
month = {oct},
title = {{Superconductivity in monolayer-trilayer phase of La$_3$Ni$_2$O$_7$ under high pressure}},
url = {http://arxiv.org/abs/2510.12250},
year = {2025}
}

@article{Shi2025a,
author = {Shi, Mengzhu and Peng, Di and Li, Yikang and Yang, Shaohua and Xing, Zhenfang and Wang, Yuzhu and Fan, Kaibao and Li, Houpu and Wu, Rongqi and Ge, Binghui and Zeng, Zhidan and Zeng, Qiaoshi and Ying, Jianjun and Wu, Tao and Chen, Xianhui},
doi = {10.1038/s41467-025-63701-x},
issn = {2041-1723},
journal = {Nature Communications},
month = {oct},
number = {1},
pages = {9141},
publisher = {Springer US},
title = {{Spin density wave rather than tetragonal structure is prerequisite for superconductivity in La$_3$Ni$_2$O$_{7-\delta}$}},
url = {https://www.nature.com/articles/s41467-025-63701-x},
volume = {16},
year = {2025}
}

@article{Li2024b,
author = {Li, Feiyu and Guo, Ning and Zheng, Qiang and Shen, Yang and Wang, Shilei and Cui, Qihui and Liu, Chao and Wang, Shanpeng and Tao, Xutang and Zhang, Guang-Ming and Zhang, Junjie},
doi = {10.1103/PhysRevMaterials.8.053401},
issn = {2475-9953},
journal = {Physical Review Materials},
month = {may},
number = {5},
pages = {053401},
title = {{Design and synthesis of three-dimensional hybrid Ruddlesden-Popper nickelate single crystals}},
url = {https://link.aps.org/doi/10.1103/PhysRevMaterials.8.053401},
volume = {8},
year = {2024}
}

@article{Inoue2026,
author = {Inoue, Daisuke and Yamakawa, Youichi and Onari, Seiichiro and Kontani, Hiroshi},
doi = {10.1038/s42005-026-02511-z},
isbn = {4200502602511},
issn = {2399-3650},
journal = {Communications Physics},
month = {feb},
number = {1},
pages = {115},
publisher = {Springer US},
title = {{Unified mechanism of charge-density-wave and high-$T_c$ superconductivity protected from oxygen vacancies in bilayer nickelates}},
url = {https://www.nature.com/articles/s42005-026-02511-z},
volume = {9},
year = {2026}
}

@article{Zhang2025a,
author = {Zhang, Mingxin and Pei, Cuiying and Peng, Di and Du, Xian and Hu, Weixiong and Cao, Yantao and Wang, Qi and Wu, Juefei and Li, Yidian and Liu, Huanyu and Wen, Chenhaoping and Song, Jing and Zhao, Yi and Li, Changhua and Cao, Weizheng and Zhu, Shihao and Zhang, Qing and Yu, Na and Cheng, Peihong and Zhang, Lili and Li, Zhiwei and Zhao, Jinkui and Chen, Yulin and Jin, Changqing and Guo, Hanjie and Wu, Congjun and Yang, Fan and Zeng, Qiaoshi and Yan, Shichao and Yang, Lexian and Qi, Yanpeng},
doi = {10.1103/PhysRevX.15.021005},
issn = {2160-3308},
journal = {Physical Review X},
month = {apr},
number = {2},
pages = {021005},
title = {{Superconductivity in Trilayer Nickelate La$_4$Ni$_3$O$_{10}$ under Pressure}},
url = {https://link.aps.org/doi/10.1103/PhysRevX.15.021005},
volume = {15},
year = {2025}
}

@article{Hou2023,
author = {Hou, Jun and Yang, Peng-Tao and Liu, Zi-Yi and Li, Jing-Yuan and Shan, Peng-Fei and Ma, Liang and Wang, Gang and Wang, Ning-Ning and Guo, Hai-Zhong and Sun, Jian-Ping and Uwatoko, Yoshiya and Wang, Meng and Zhang, Guang-Ming and Wang, Bo-Sen and Cheng, Jin-Guang},
doi = {10.1088/0256-307X/40/11/117302},
issn = {0256-307X},
journal = {Chinese Physics Letters},
month = {oct},
number = {11},
pages = {117302},
title = {{Emergence of High-Temperature Superconducting Phase in Pressurized La$_3$Ni$_2$O$_7$ Crystals}},
url = {https://iopscience.iop.org/article/10.1088/0256-307X/40/11/117302},
volume = {40},
year = {2023}
}

@article{Wang2023a,
author = {Wang, Luhong and Li, Yan and Xie, Shengyi and Liu, Fuyang and Sun, Hualei and Huang, Chaoxin and Gao, Yang and Nakagawa, Takeshi and Fu, Boyang and Dong, Bo and Cao, Zhenhui and Yu, Runze and Kawaguchi, Saori I. and Kadobayashi, Hirokazu and Wang, Meng and Jin, Changqing and Mao, Ho-kwang and Liu, Haozhe},
doi = {10.1021/jacs.3c13094},
issn = {0002-7863},
journal = {Journal of the American Chemical Society},
month = {mar},
number = {11},
pages = {7506--7514},
title = {{Structure Responsible for the Superconducting State in La$_3$Ni$_2$O$_7$ at High-Pressure and Low-Temperature Conditions}},
url = {https://pubs.acs.org/doi/10.1021/jacs.3c13094},
volume = {146},
year = {2024}
}

@article{Puphal2023,
author = {Puphal, Pascal and Reiss, Pascal and Enderlein, Niklas and Wu, Y.-M. and Khaliullin, Giniyat and Sundaramurthy, Vignesh and Priessnitz, Tim and Knauft, Manuel and Suthar, A. and Richter, Lea and Isobe, Masahiko and van Aken, P. A. and Takagi, Hidenori and Keimer, Bernhard and Suyolcu, Y. E. and Wehinger, Bj{\"{o}}rn and Hansmann, Philipp and Hepting, Matthias},
doi = {10.1103/PhysRevLett.133.146002},
issn = {0031-9007},
journal = {Physical Review Letters},
month = {oct},
number = {14},
pages = {146002},
title = {{Unconventional Crystal Structure of the High-Pressure Superconductor La$_3$Ni$_2$O$_7$}},
url = {https://link.aps.org/doi/10.1103/PhysRevLett.133.146002},
volume = {133},
year = {2024}
}

@article{Wang2023,
author = {Wang, Haozhe and Chen, Long and Rutherford, Aya and Zhou, Haidong and Xie, Weiwei},
doi = {10.1021/acs.inorgchem.3c04474},
issn = {0020-1669},
journal = {Inorganic Chemistry},
month = {mar},
number = {11},
pages = {5020--5026},
title = {{Long-Range Structural Order in a Hidden Phase of Ruddlesden-Popper Bilayer Nickelate La$_3$Ni$_2$O$_7$}},
url = {https://pubs.acs.org/doi/10.1021/acs.inorgchem.3c04474},
volume = {63},
year = {2024}
}

@article{Li2024,
author = {Li, Qing and Zhang, Ying-Jie and Xiang, Zhe-Ning and Zhang, Yuhang and Zhu, Xiyu and Wen, Hai-Hu},
doi = {10.1088/0256-307X/41/1/017401},
issn = {0256-307X},
journal = {Chinese Physics Letters},
month = {jan},
number = {1},
pages = {017401},
title = {{Signature of Superconductivity in Pressurized La$_4$Ni$_3$O$_{10}$}},
url = {https://iopscience.iop.org/article/10.1088/0256-307X/41/1/017401},
volume = {41},
year = {2024}
}

@article{Zhou2025,
author = {Zhou, Yazhou and Guo, Jing and Cai, Shu and Sun, Hualei and Li, Chengyu and Zhao, Jinyu and Wang, Pengyu and Han, Jinyu and Chen, Xintian and Chen, Yongjin and Wu, Qi and Ding, Yang and Xiang, Tao and Mao, Ho-kwang and Sun, Liling},
doi = {10.1063/5.0247684},
issn = {2468-2047},
journal = {Matter and Radiation at Extremes},
month = {mar},
number = {2},
title = {{Investigations of key issues on the reproducibility of high-$T_c$ superconductivity emerging from compressed La$_3$Ni$_2$O$_7$}},
url = {https://pubs.aip.org/mre/article/10/2/027801/3331819/Investigations-of-key-issues-on-the},
volume = {10},
year = {2025}
}

@article{Wang2024,
author = {Wang, Ningning and Wang, Gang and Shen, Xiaoling and Hou, Jun and Luo, Jun and Ma, Xiaoping and Yang, Huaixin and Shi, Lifen and Dou, Jie and Feng, Jie and Yang, Jie and Shi, Yunqing and Ren, Zhian and Ma, Hanming and Yang, Pengtao and Liu, Ziyi and Liu, Yue and Zhang, Hua and Dong, Xiaoli and Wang, Yuxin and Jiang, Kun and Hu, Jiangping and Nagasaki, Shoko and Kitagawa, Kentaro and Calder, Stuart and Yan, Jiaqiang and Sun, Jianping and Wang, Bosen and Zhou, Rui and Uwatoko, Yoshiya and Cheng, Jinguang},
doi = {10.1038/s41586-024-07996-8},
issn = {0028-0836},
journal = {Nature},
month = {oct},
number = {8034},
pages = {579--584},
pmid = {39358510},
publisher = {Springer US},
title = {{Bulk high-temperature superconductivity in pressurized tetragonal La$_2$PrNi$_2$O$_7$}},
url = {https://www.nature.com/articles/s41586-024-07996-8},
volume = {634},
year = {2024}
}

@article{Khasanov2024,
author = {Khasanov, Rustem and Hicken, Thomas J. and Gawryluk, Dariusz J. and Sazgari, Vahid and Plokhikh, Igor and Sorel, Lo{\"{i}}c Pierre and Bartkowiak, Marek and B{\"{o}}tzel, Steffen and Lechermann, Frank and Eremin, Ilya M. and Luetkens, Hubertus and Guguchia, Zurab},
doi = {10.1038/s41567-024-02754-z},
issn = {1745-2473},
journal = {Nature Physics},
month = {mar},
number = {3},
pages = {430--436},
title = {{Pressure-enhanced splitting of density wave transitions in La$_3$Ni$_2$O$_{7-\delta}$}},
url = {https://www.nature.com/articles/s41567-024-02754-z},
volume = {21},
year = {2025}
}

@article{Chen2025,
author = {Chen, Yu and Zhang, Ke and Xu, Minghui and Zhao, Yan and Xiao, Haiyan and Qiao, Liang},
doi = {10.1007/s11433-024-2556-4},
issn = {1674-7348},
journal = {Science China Physics, Mechanics \& Astronomy},
month = {apr},
number = {4},
pages = {247411},
title = {{Oxygen deficiency mechanism of La$_3$Ni$_2$O$_{7-\delta}$ under pressure}},
url = {https://link.springer.com/10.1007/s11433-024-2556-4},
volume = {68},
year = {2025}
}

@article{Jiang2025a,
archivePrefix = {arXiv},
arxivId = {2512.00301},
journal = {arXiv (online)},
author = {Jiang, Peiheng and Li, Jie and Cao, Yu-Han and Cao, Xiaodong and Zhong, Zhicheng and Lu, Yi and Wang, Qiang-Hua},
eprint = {2512.00301},
title = {{Dual instability of superconductivity from oxygen defects in La$_3$Ni$_2$O$_{7+\delta}$}},
url = {http://arxiv.org/abs/2512.00301},
year = {2025}
}

@article{Hu2026,
author = {Hu, Xin-Ran and Yu, Jia and Yi, Jun-Kun and Gu, Ya-Dong and Feng, Jun-Jie and Sun, Xie-Yu and Liu, Qing-Song and Shi, Yun-Qing and Zhou, Meng-Hu and Sun, Hui-Ran and Jiang, Chang-Sheng and Huang, Qi-Heng and Han, Tao and Long, Ming-Sheng and Hou, Xing-Yuan and Ge, Bing-Hui and Song, Dong-Sheng and Ren, Zhi-An and Mu, Qing-Ge and Shan, Lei},
doi = {10.1088/1674-1056/ae5050},
issn = {1674-1056},
journal = {Chinese Physics B},
month = {may},
number = {5},
pages = {057407},
title = {{Dual roles of oxygen-annealing in modulating the superconductivity of La$_3$Ni$_2$O$_{7+\delta}$}},
url = {https://iopscience.iop.org/article/10.1088/1674-1056/ae5050},
volume = {35},
year = {2026}
}

@article{Dong2024,
author = {Dong, Zehao and Huo, Mengwu and Li, Jie and Li, Jingyuan and Li, Pengcheng and Sun, Hualei and Gu, Lin and Lu, Yi and Wang, Meng and Wang, Yayu and Chen, Zhen},
doi = {10.1038/s41586-024-07482-1},
issn = {0028-0836},
journal = {Nature},
month = {jun},
number = {8018},
pages = {847--852},
pmid = {38839959},
title = {{Visualization of oxygen vacancies and self-doped ligand holes in La$_3$Ni$_2$O$_{7-\delta}$}},
url = {https://www.nature.com/articles/s41586-024-07482-1},
volume = {630},
year = {2024}
}

@article{Misawa2026,
archivePrefix = {arXiv},
arxivId = {2603.25119},
author = {Misawa, Ryo and Kitou, Shunsuke and Sun, Jian-Ping and Yu, Yingpeng and Koyama, Chihaya and Nakamura, Yuiga and Arima, Taka-hisa and Cheng, Jin-Guang and Hirschberger, Max},
eprint = {2603.25119},
journal = {arXiv (online)},
month = {mar},
title = {{Polar, checkerboard charge order in bilayer nickelate La$_3$Ni$_2$O$_7$}},
url = {http://arxiv.org/abs/2603.25119},
year = {2026}
}

@article{Zhang2024,
author = {Zhang, Yanan and Su, Dajun and Huang, Yanen and Shan, Zhaoyang and Sun, Hualei and Huo, Mengwu and Ye, Kaixin and Zhang, Jiawen and Yang, Zihan and Xu, Yongkang and Su, Yi and Li, Rui and Smidman, Michael and Wang, Meng and Jiao, Lin and Yuan, Huiqiu},
doi = {10.1038/s41567-024-02515-y},
issn = {1745-2473},
journal = {Nature Physics},
month = {aug},
number = {8},
pages = {1269--1273},
title = {{High-temperature superconductivity with zero resistance and strange-metal behaviour in La$_3$Ni$_2$O$_{7-\delta}$}},
url = {https://www.nature.com/articles/s41567-024-02515-y},
volume = {20},
year = {2024}
}

@article{Chen2024,
author = {Chen, Kaiwen and Liu, Xiangqi and Jiao, Jiachen and Zou, Muyuan and Jiang, Chengyu and Li, Xin and Luo, Yixuan and Wu, Qiong and Zhang, Ningyuan and Guo, Yanfeng and Shu, Lei},
doi = {10.1103/PhysRevLett.132.256503},
issn = {0031-9007},
journal = {Physical Review Letters},
month = {jun},
number = {25},
pages = {256503},
pmid = {38996236},
publisher = {American Physical Society},
title = {{Evidence of Spin Density Waves in La$_3$Ni$_2$O$_{7-\delta}$}},
url = {https://link.aps.org/doi/10.1103/PhysRevLett.132.256503},
volume = {132},
year = {2024}
}

@article{Samanta2025,
author = {Samanta, Bipasa and Georgescu, Alexandru B.},
doi = {10.1103/r5yy-8m64},
issn = {2475-9953},
journal = {Physical Review Materials},
month = {nov},
number = {11},
pages = {114801},
publisher = {American Physical Society},
title = {{In-plane Ni-O-Ni bond angles as structural fingerprints of superconductivity in layered nickelates: Effects of pressure, strain, layering, and correlations}},
url = {https://link.aps.org/doi/10.1103/r5yy-8m64},
volume = {9},
year = {2025}
}

@article{Sakurai2026,
author = {Sakurai, Hiroya and Takano, Yoshihiko},
doi = {10.1088/1361-648X/ae4155},
issn = {0953-8984},
journal = {Journal of Physics: Condensed Matter},
month = {feb},
number = {7},
pages = {073002},
pmid = {41633048},
title = {{Superconducting lanthanum nickel oxides with bilayered and trilayered crystal structures}},
url = {https://iopscience.iop.org/article/10.1088/1361-648X/ae4155},
volume = {38},
year = {2026}
}

@article{Wang2023b,
author = {Wang, G. and Wang, N. N. and Shen, X. L. and Hou, Jun and Ma, L. and Shi, L. F. and Ren, Z. A. and Gu, Y. D. and Ma, H. M. and Yang, P. T. and Liu, Z. Y. and Guo, H. Z. and Sun, J. P. and Zhang, G. M. and Calder, S. and Yan, J.-Q. and Wang, B. S. and Uwatoko, Yoshiya and Cheng, J.-G.},
doi = {10.1103/PhysRevX.14.011040},
issn = {2160-3308},
journal = {Physical Review X},
month = {mar},
number = {1},
pages = {011040},
title = {{Pressure-Induced Superconductivity In Polycrystalline La$_3$Ni$_2$O$_{7-\delta}$}},
url = {https://link.aps.org/doi/10.1103/PhysRevX.14.011040},
volume = {14},
year = {2024}
}

@article{Chen2024a,
author = {Chen, Xinglong and Zhang, Junjie and Thind, Arashdeep S. and Sharma, Shekhar and LaBollita, Harrison and Peterson, Gordon and Zheng, Hong and Phelan, Daniel P. and Botana, Antia S. and Klie, Robert F. and Mitchell, J. F.},
doi = {10.1021/jacs.3c14052},
issn = {0002-7863},
journal = {Journal of the American Chemical Society},
month = {feb},
number = {6},
pages = {3640--3645},
title = {{Polymorphism in the Ruddlesden–Popper Nickelate La$_3$Ni$_2$O$_7$: Discovery of a Hidden Phase with Distinctive Layer Stacking}},
url = {https://pubs.acs.org/doi/10.1021/jacs.3c14052},
volume = {146},
year = {2024}
}

@article{Liu2023,
author = {Liu, Zengjia and Sun, Hualei and Huo, Mengwu and Ma, Xiaoyan and Ji, Yi and Yi, Enkui and Li, Lisi and Liu, Hui and Yu, Jia and Zhang, Ziyou and Chen, Zhiqiang and Liang, Feixiang and Dong, Hongliang and Guo, Hanjie and Zhong, Dingyong and Shen, Bing and Li, Shiliang and Wang, Meng},
doi = {10.1007/s11433-022-1962-4},
issn = {1674-7348},
journal = {Science China Physics, Mechanics \& Astronomy},
month = {jan},
number = {1},
pages = {217411},
title = {{Evidence for charge and spin density waves in single crystals of La$_3$Ni$_2$O$_7$ and La$_3$Ni$_2$O$_6$}},
url = {https://link.springer.com/10.1007/s11433-022-1962-4},
volume = {66},
year = {2023}
}

@article{Sakakibara2024,
author = {Sakakibara, Hirofumi and Ochi, Masayuki and Nagata, Hibiki and Ueki, Yuta and Sakurai, Hiroya and Matsumoto, Ryo and Terashima, Kensei and Hirose, Keisuke and Ohta, Hiroto and Kato, Masaki and Takano, Yoshihiko and Kuroki, Kazuhiko},
doi = {10.1103/PhysRevB.109.144511},
issn = {2469-9950},
journal = {Physical Review B},
month = {apr},
number = {14},
pages = {144511},
publisher = {American Physical Society},
title = {{Theoretical analysis on the possibility of superconductivity in the trilayer Ruddlesden-Popper nickelate La$_4$Ni$_3$O$_{10}$ under pressure and its experimental examination: Comparison with La$_3$Ni$_2$O$_7$}},
url = {https://link.aps.org/doi/10.1103/PhysRevB.109.144511},
volume = {109},
year = {2024}
}

@article{Khasanov2026,
author = {Khasanov, Rustem and Sazgari, Vahid and Hicken, Thomas J and Plokhikh, Igor and Medarde, Marisa and Pomjakushina, Ekaterina and Keller, Lukas and Pomjakushin, Vladimir and Bartkowiak, Marek and Kr{\'{o}}lak, Szymon and Winiarski, Micha{\l} J and Steppke, Alexander and Krieger, Jonas A and Luetkens, Hubertus and Klimczuk, Tomasz and Schneider, Christof W and Gawryluk, Dariusz J and Guguchia, Zurab},
doi = {10.1103/nrqn-m22c},
journal = {Physical Review Research},
number = {01},
pages = {013249},
publisher = {American Physical Society},
title = {{Pressure and oxygen-isotope substitution on density-wave transitions in La$_4$Ni$_3$O$_{10}$}},
url = {https://journals.aps.org/prresearch/abstract/10.1103/nrqn-m22c},
volume = {8},
year = {2026}
}

@article{Yuan2024,
author = {Yuan, Ning and Elghandour, Ahmed and Arneth, Jan and Dey, Kaustav and Klingeler, Rüdiger},
journal = {Journal of Crystal Growth},
title = {{High-pressure crystal growth and investigation of the metal-to-metal transition of Ruddlesden–Popper trilayer nickelates La$_4$Ni$_3$O$_{10}$}},
year = {2024},
issn = {0022-0248},
month = Feb,
pages = {127511},
volume = {627},
doi = {10.1016/j.jcrysgro.2023.127511},
publisher = {Elsevier BV},
}

@article{Keimer2015,
author = {Keimer, B. and Kivelson, S. A. and Norman, M. R. and Uchida, S. and Zaanen, J.},
doi = {10.1038/nature14165},
issn = {0028-0836},
journal = {Nature},
month = {feb},
number = {7538},
pages = {179--186},
publisher = {Nature Publishing Group},
title = {{From quantum matter to high-temperature superconductivity in copper oxides}},
url = {http://www.nature.com/articles/nature14165},
volume = {518},
year = {2015}
}

@article{Matsuoka2018,
author = {Matsuoka, H and Nakano, M and Uchida, M and Kawasaki, M and Iwasa, Y},
doi = {10.1103/PhysRevB.98.144506},
journal = {Physical Review B},
number = {14},
pages = {144506},
publisher = {American Physical Society},
title = {{Signatures of charge-order correlations in transport properties of electron-doped cuprate superconductors}},
volume = {98},
year = {2018},
url = {https://journals.aps.org/prb/abstract/10.1103/PhysRevB.98.144506}
}

@article{Lee2021,
author = {Lee, W S and Zhou, Ke-jin and Hepting, M and Li, J and Nag, A and Walters, A C and Garcia-Fernandez, M and Robarts, H C and Hashimoto, M and Lu, H and Nosarzewski, B and Song, D and Eisaki, H and Shen, Z X and Moritz, B and Zaanen, J and Devereaux, T P},
doi = {10.1038/s41567-020-0993-7},
journal = {Nature Physics},
pages = {53--57},
title = {{Spectroscopic fingerprint of charge order melting driven by quantum fluctuations in a cuprate}},
volume = {17},
year = {2021},
url = {https://www.nature.com/articles/s41567-020-0993-7}
}

\end{document}